\documentstyle[12pt,aasms4]{article}

\slugcomment{}

\lefthead{Enya et al.}
\righthead{$JHK'$ Imaging  Photometry of AGNs}

\begin{document}

\title{$JHK'$ Imaging Photometry of Seyfert 1 AGNs and Quasars I: \\
            Multi-Aperture Photometry}

\author{Keigo Enya\altaffilmark{1}, Yuzuru Yoshii\altaffilmark{1,4},
   Yukiyasu Kobayashi\altaffilmark{2}, Takeo Minezaki\altaffilmark{1}, 
   Masahiro Suganuma\altaffilmark{3}, Hiroyuki Tomita\altaffilmark{3}
   and Bruce A. Peterson\altaffilmark{5}}

\vspace{20mm}

\affil{$^1$ Institute of Astronomy, School of Science,
              University of Tokyo, Osawa 2-21-1,  
              Mitaka, Tokyo 181-8588, Japan}
\affil{$^2$ National Astronomical Observatory,
              Osawa 2-21-1, Mitaka, Tokyo 181-8588, Japan}
\affil{$^3$ Department of Astronomy, University of Tokyo,
             Hongo 7-3-1, Bunkyo-ku, Tokyo  113-0033, Japan}
\affil{$^4$ Research Center for the Early Universe (RESCEU),
          School of Science, University of Tokyo,
             Hongo 7-3-1, Bunkyo-ku, Tokyo 113-0033, Japan}
\affil{$^5$ Research School of Astronomy and Astrophysics, The
         Australian National University, Weston Creek, ACT 2611,
         Australia}

%%%%%%%%%%%%%%%%%%%%%%%%%%%%%%%%%%%%%%%%%%%%%%%%%%%%%%%%%%%%%%%%%%%%%%%%%

\begin{abstract}

Near-infrared $JHK'$ imaging photometry was obtained of  331 AGNs consisting 
mainly of Seyfert 1 AGNs and quasars (QSOs).   This sample was selected to 
cover a range of radio emission strength, redshift  from $z=0$ to 1, 
and absolute $B$-magnitude from $M_B=-29$ mag to $-18$ mag.  Among 
low-$z$ AGNs with $z<0.3$, Seyfert $1-1.5$ AGNs are distributed over a region
from a location
typical of ``galaxies'' to a location typical of ``QSOs'' in 
the two-color $J-H$ to $H-K'$ diagram, but Seyfert $1.8-2$ AGNs are 
distributed around the location of ``galaxies''.  Moreover, bright AGNs with
respect to absolute $B$-magnitude are distributed near the location of 
``QSOs'', 
while faint AGNs are near the location of ``galaxies''.  The distribution 
of such 
low-$z$ AGNs in this diagram was found to have little dependence  on their 
6 cm 
radio flux.  The near-infrared colors of the AGNs observed with an aperture of
7 pixels ($7.49''$) are more QSO-like than those observed 
with larger apertures up to 15 pixels ($16.1''$). 
This aperture effect may  be explained
by contamination from the light of host galaxies 
within larger apertures.  
This effect is more prominent for less luminous AGNs.

\end{abstract}


%%%%%%%%%%%%%%%%%%%%%%%%%%%%%%%%%%%%%%%%%%%%%%%%%%%%%%%%%%%%%%%%%%%%%%%%%%%%%%
\section{Introduction}

Near-infrared (NIR) observations are  useful to study the dust distribution
in Active Galactic Nuclei (AGNs).  The unified model assumes  a dust torus 
around the central engine in which  the torus viewed at different 
angles from the line of sight explains the difference between Seyfert 1 
AGNs and Seyfert 2 AGNs (Antonucci \& Miller 1985). 

NIR observations are also useful to derive general features in the 
spectral energy distributions (SEDs) of AGNs, and to examin how much these 
SEDs are contaminated by the light from host galaxies.  Previous authors 
(Sanders et al. 1989; Neugebuer et al. 1979) presented the SEDs of 109 
AGNs which show that the  general features of a 1$\mu m$ minimum 
and a 3$\mu m$ bump were present in the SEDs of many AGNs.  
Kobayashi et al. (1993) obtained the SEDs of 
14 quasars by 16 channel prism spectrophotometry between $0.95\mu m$ 
and $2.5\mu m$, and found that all the SEDs are characterized as having 
a black body SED with a typical temperature of 1500K correspondng to
that of dust sublimation, 
seperated by a power-law component with a variety of power indicies.  
Elvis et al. (1994) presented the SEDs of 47 AGNs (29 radio-quiet and 
18 radio-loud AGNs) over a wide range of wavelength from the X-ray to 
the radio region.  

Hunt et al. (1997) obtained  $JHK$ imags of 26 nearby AGNs.  Alonso-Herrero, 
Ward \& Kotilainen (1996) obtained  the $JHKL$ imags of 13 Seyfert 2 AGNs and 
decomposed their SEDs into the stellar and non-stellar components.
Because NIR imaging photometry  has been limited to nearby AGNs, the sample 
size 
has also been limited.  Motivated by the  need to expand the sample size at 
least by an order of magnitude, we undertook a program of imaging a few 
hundred 
AGNs in the NIR region, carried out  a statistical study AGN SEDs with 
the largest sample ever made.  Many of them were observed on two  different 
nights seperated by a year or more, for the purpose of detecting the NIR 
variability of the AGNs in our sample. 

In a series of three papers, we give our  results from three-years 
of observations.  In this Paper I, we present NIR magnitudes of more than 
300 AGNs derived by multi-color, multi-aperture imaging photometory.  
Analysis and discussion of AGN variability will be presented in the 
forthcoming Papers II and III.  
We are currently conducting a project called MAGNUM (an acronym of 
Multicolor Active Galactic NUclei Monitoring, Kobayashi et al. 1998a, 1998b) 
which monitors AGN
in the 11 passbands of $UBVRIZJHKL$, and aims to determine the 
distances to many AGNs by measurements of the delay time between optical and
NIR variabilities.  Therefore, this paper is not only an individual study, 
but also a preliminary study to select target AGNs for the MAGNUM Project.

%%%%%%%%%%%%%%%%%%%%%%%%%%%%%%%%%%%%%%%%%%%%%%%%%%%%%%%%%%%%%%%%%%%%%%%%%%%%%%
\section{Sample Selection and Observations}

%%%%%%%%%%%%%%%%%%%%%%%%%%%%%%%%%
\subsection{sample selection}
%%%%%%%%%%%%%%%%%%%%%%%%%%%%%%%%%

In the beginning of this study, all AGNs were selected from the 6th version 
of the Quasars and Active Galactic Nuclei catalog (hereafter referred to 
the VV catalog; Veron-Cetty and Veron 1993).  Additional AGNs were
selected from new versions that were released 
during this sudy, the 7th version (Veron-Cetty and Veron 1996) and the 8th 
version (Veron-Cetty and Veron 1998).  We summarize the selection criteria 
such as (1) coordinates, 
(2) AGN types, (3) absolute $B$-magnitudes, and (4) redshifts:

\noindent
(1)Declinations were selected from $\delta=-10$ degrees to $+50$ degrees, 
allowing for a wide coverage of right ascension, $\alpha$. This is necessary 
in order for the MAGNUM Project to observe many AGNs under good conditions 
during 
the entire year from Haleakala, on the Hawaiian Island of Maui
at a latitude of $+20$ degrees where the MAGNUM Observatory is situated.
      
\noindent
(2)Seyfert 1 AGNs and quasars were selected, excluding Seyfert 2 AGNs 
and BL Lac objects. This is necessary  in order for the MAGNUM Project to 
observe 
thermal radiation from the innermost region of the dust torus which surrounds 
the central engine of AGNs.  We excluded Seyfert 2 AGNs, because they are 
aligned with the dust torus edge-on so that the inner region is obscured.  
We also excluded BL Lac objects, because their SEDs are known to exhibit 
only weak thermal radiation from hot dust.

\noindent
(3)Absolute $B$-magnitudes were selected to span from $M_B=-29$ to 
$-18$, enabling a statistical study of the $M_B$-dependence of various other 
spectral features of AGNs.  This is necessary in order to discuss a 
Malmquist-type 
bias which affects an interpretation of any statistical study from bright, 
distant AGNs in a sample. 

\noindent
(4)Redshifts were selected to span from $z=0$ to 1, enabling a statistical 
study of $z$-dependence of various other spectral features of AGNs.  
By considering that the maximum wavelength covered by the MAGNUM 
camera is the $L$ band,  
the redshifts were limited to below unity. Otherwise the thermal radiation 
peaked at $2\micron $ corresponding to the 1500K temperature of dust 
sublimation shifts to much longer wavelengths, beyond the $L$-band filter.

Table 1 tabulates the basic quantities of 331 AGNs selected in this 
study. For the purpose of illustration, Fig. 1 shows the distribution of 
$\alpha$ and $\delta$ for all AGNs in the sample, and Fig. 2 shows their 
distribution of $M_B$ and $z$.

%%%%%%%%%%%%%%%%%%%%%%%%%%
\subsection{observation}
%%%%%%%%%%%%%%%%%%%%%%%%%%

All AGNs and quasars in the sample were observed with the 1.3m infrared 
telescope at the Institute of Space and Astronautical Science (ISAS), 
Japan. We used the PICNIC camera developed at the National Astronomical 
Observatory, Japan (NAOJ) for multi-color broard-band imaging (Kobayashi 
et al. 1994).  The PICNIC camera has a NICMOS3 array of $256 \times 256$ 
pixels, corresponding to a field of view of $4.57' \times 4.57'$ and a pixel 
scale of $1.07''$ pixel$^{-1}$.  Imaging photometry was done in the NIR 
broad bands with the $J$, $H$, and $K'$ filters.  In order to decrease 
the thermal sky background, we used $K'$ filter which has the same 
transmission curve as the 2MASS $K_s$ filter (McLeod et al. 1995).

Our observational runs consist of three periods (January 1996$-$April 
1996, November 1996$-$February 1997, and December 1997$-$April 1998).
More than 300 AGNs were observed in the first and second periods.
In the third period, however, more than 200 AGNs that had been 
observed in previous two periods were again observed in order to 
determine their variability.  Analysis and discussion 
of the variability of AGNs will be 
presented elsewhere (Papers II and III). 

The AGNs and quasars were imaged in the $J$, $H$, and $K'$ bands 
by stepping the telescope in a raster pattern.
The tipical exposure time of  each 
single frame was 35 sec ($J$), 17 sec ($H$), and 8 sec ($K'$), while 
a shorter exposure time was used if saturation might occur because of 
either a high thermal background, or a large flux from an object in
the frame. The minimum 
number of frames for one object was 4 frames with $2\times 2$ positions 
($J$), 4 frames with $2\times 2$ positions ($H$), and 9 frames with 
$3\times 3$ positions ($K'$).  More frames were obtained for fainter 
objects.  The maximum number of frames for one object was 50 frames ($J$), 
200 frames ($H$), and 200 frames ($K'$), with $5 \times 5$ positions in 
common.

Two photometric standard stars with different elevations were observed 
three times in one night, that is, at the beginning of observations, 
midnight and at the end of observations. These standard stars were imaged 
with a $3 \times 3$ raster,  and two frames were obtained at each 
position. In this way a total of 18 frames were obtained for one star,
and the acquisition of all the frames in the $J$, $H$, and $K'$ bands took 
about 15 minutes per star.  If such schedule was impossible because of 
bad weather or other reasons, the standard stars were observed before and 
after 
the observations of the AGNs.  

Each night after the AGNs and standard stars were observed, dome flat 
images were obtained by turning a calibration lamp on and off in front 
of white board, then the dark current was measured with a cold opaque 
shutter blocking all external radiation. 

%%%%%%%%%%%%%%%%%%%%%%%%%%%%%%%%%%%%%%%%%%%%%%%%%%%%%%%%%%%%%%%%%%%%%%%%%%%%%%
\section{Reduction}

%%%%%%%%%%%%%%%%%%%%%%%%%%%%%%%%%%%%%%%%%%%%%%%%%%%
\subsection{image reduction}\label{sec_image_red}
%%%%%%%%%%%%%%%%%%%%%%%%%%%%%%%%%%%%%%%%%%%%%%%%%%%

All the frames of AGNs and standard 
stars were obtained with short integration times and with dithering, 
which produced a large number of the frames in the end.  A short 
integration time was used to avoid  saturation  
during measurements with high sky background.  The dithering was used 
to minimize the effect of bad pixels and to make a sky flat 
of high quality.  

The software system specialized to analyze the data obtained by the 
PICNIC camera (hereafter PICRED) was used for the reduction.  
PICRED is a semi-automated system, requiring 
manual operations and human decisions in order to deal with various 
types of data, from star formong regins to quasars. 
For  our case of reducing an enormous 
amount of the data by repeating much the same procedure, the software 
system was made fully automated.

%%%%%%%%%%%%%%%%%%%%%%%%%%%%%%%%%%%%%%%%
\subsection{photometric calibration}
%%%%%%%%%%%%%%%%%%%%%%%%%%%%%%%%%%%%%%%%

Each time a pair of standard stars was observed, 18 frames were 
taken first for one star at high elevation for each passband, in order, 
from $J$ to $H$ and then to $K'$,  and then similarly for another star 
at low elevation.  This procedure was repeated three times in one night, 
mostly for different pairs. Abnormal data, deviating remarkably 
from others, would possibly occur due to major three factors, such as 
obscuration by thin clouds that were not detected during the observation, 
the effect of bad pixels, or misidentification of the target.  
 
In the beginning of reduction process, bad frames, if any, were 
excluded, and only the remaining frames were used to determine the 
instrumental magnitudes from which the median $m({\rm inst})$ and 
dispersion $\sigma_d$ were obtained for each star in each passband.  
The median $m({\rm inst})$ was then transformed to the already 
calibrated(known) magnitude of each star, by taking a linear fit in a plot 
of $\Delta m=m({\rm inst})-m({\rm calib})$ against airmass. 
The error $\sigma_a$ in transformation was also estimated.  In this way, 
we determined the aperture $J$, $H$, and $K'$ magnitudes of AGNs 
using four different apertures of 7, 10, 12, and 15 pixels in radius. 
The total error in magnitude is given by $\sigma_m^2=\sigma_d^2+\sigma_a^2$.  
These aperture magnitudes and errors are tabulated in Tables 2a 
($J$ band), 2b ($H$ band) and 2c($K'$ band).

%%%%%%%%%%%%%%%%%%%%%%%%%%%%%%%%%%%%%%%%%%%%%%%%%%%%
%\subsection{error related to atmospheric variation}
%%%%%%%%%%%%%%%%%%%%%%%%%%%%%%%%%%%%%%%%%%%%%%%%%%%%

Figure 3  shows the frequency distribution of 
$\sigma_a$ for the aperture of 15 pixels in the $J$, $H$, and $K'$ bands.  
For this large aperture of 15 pixels, the error $\sigma_a$ 
would originate from the variation of atmospheric transmissivity 
rather than the variation in the  seeing during the night.   
The passband of longer wavelength has the smaller 
$\sigma_a$ distribution.  
This feature indicates that 
detected photon counts in the $J$ 
band as compared to the $H$ and $K'$ bands
is more sensitive to the variation of atmosphere 
transmissivity.

%%%%%%%%%%%%%%%%%%%%%%
%\subsection{error related to dithering}\label{subsec_err_dith}
%%%%%%%%%%%%%%%%%%%%%%

Figure 4 shows the frequency distribution of 
$\sigma_d$ for the aperture of 15 pixels in the $J$, $H$, and $K'$ bands.  
In each panel, the distributions shown are  based on our observations 
in the first and third periods ({\it solid line}; January 1996$-$April 
1996, December 1997$-$April 1998) and in the second period ({\it dashed 
line}; November 1996$-$February 1997).  These distributions are similar 
to each other, except that the peak for the first and third periods 
occurs at larger $\sigma_d$ than that for the second period. 

Irrespective of the observational period, however, the dispersion 
$\sigma_d$ is larger than that expected from the high S/N ratio ($\leq 0.01$ 
mag), or from small changes of atmospheric transmissivity and airmass 
during the short exposure time of about 15 minutes.  Therefore, the 
dispersion $\sigma_d$ may originate from systematic errors in flat fielding.

Our observations in the second period were made by avoiding use of the 
fourth quadrant of the detector because it 
was out of order.  The dithering shifts in the second period 
were at most about 30 arcsec, only  half of those for 
normal observations in the first and third periods.  The resulting 
difference in dithering patterns gives a measure of the flat fielding error 
over some tens of arcsec that is comprable to $\sigma_d$.  

The average and median of $\sigma_d$ for the first and third periods 
are 0.033 mag and 0.027 mag, respectively. Corrresponding values
for the second period are smaller.  
This result is almost independent of passband, 
in contrast to that for $\sigma_a$.   Thus, $\sigma_a$ is larger than 
$\sigma_d$ in the $J$ band, while the converse is true in the $K'$ 
band.

%%%%%%%%%%%%%%%%%%%%%%%%%%%%%%%%%%%%%%%%%%%%%%%%%%%%%%%%%%%%%%%%%%%%%%%%%%%%%%
\section{Results}

The ratio of radio flux  $f_{\nu}({\rm 6cm})$ relative to optical 
$V$-band flux $f_{\nu}(V)$ is used as a measure of radio strength of the 
AGNs in our sample.  Figure 5 shows the distribution of 
this ratio based on the data taken from the VV catalog.  The values of
this ratio range over several orders, but are localized around 1 and 
1000.  Here, in this paper, the AGNs with $f_{\nu}({\rm 6cm})/f_{\nu}(V) 
<10$ and no radio detection are classified as radio quiet, and those with 
$f_{\nu}({\rm 6cm})/f_{\nu}(V) > 100$ as radio loud.  

It is known that there are two typical locations in the two-color $J-H$ 
to $H-K'$ diagram, such as ($J-H$, $H-K'$)=(0.8, 1) for the ``QSOs'' (Hyland
\& Allen 1982), and (0.7, 0.3) for ``galaxies'' (Willer et al. 1984).  
AGNs in our sample are found to be 
distributed in a wide region from ``QSOs'' to ``galaxies''.  
In fact, brighter AGNs tend to populate the diagram 
near the ``QSOs'', while fainter 
AGNs or Syfert $1.8-2$ AGNs near the ``galaxies''.  No such localization, 
however, occurs if the sample is divided into the radio-quiet and 
radio-loud AGNs.
 
All these features are more clearly seen in Fig. 6, where averages of 
$J-H$ and $H-K'$ colors and their errors for low-$z$ AGNs with $0<z<0.3$ 
are plotted as a function of $M_B$, Seyfert type, and 
$f_{\nu}({\rm 6cm})/f_{\nu}(V)$.
Shown are the results for the four different apertures of 7, 10, 12, and 
15 pixels.  Here, averages are taken with no weights, and errorbars are 
the standard deviations of the colors for an aperture of  
7 pixels. 

The $H-K'$ color becomes monotonically bluer from $M_B=-27$ to $-21$, 
irrespective of aperture.  AGNs with $M_B<-25$ have $H-K'\sim 1$ and 
are QSO-like, while those with $M_B>-22$ have  $H-K'\sim 0.5$ and are 
galaxy-like.  On the other hand, the $J-H$ color stays at about 0.8, 
and its $M_B$-dependence is much weaker than that for $H-K'$.

Furthermore, for AGNs with $M_B>-24$, the $H-K'$ color 
determined with larger aperture is more galaxy-like.  
This feature originats from  the color gradient in the central 
region where the AGN dominates to the outer region where 
the host galaxy dominates.  It should be noticed that
the blueward color shift with the use of larger aperture is more 
significant for fainter AGNs with $M_B>-24$ and for the $H-K'$ color 
rather than for $J-H$.  This feature can be explained naturally by 
considering that the host galaxy becomes more visible within larger aperture
and its SED is enhanced over the AGN at NIR wavelengths in rest frame.  

The monotonical trend in the average colors is also seen by changing Seyfert 
type from 1 to 2. Although Seyfert $1-1.5$ AGNs are in between 
``QSOs'' and ``galaxies'', Seyfert $1.8-2$ AGNs are galaxy-like. This 
partly reflects the tendency that Seyfert $1.8-2$ AGNs are, on the whole, 
faint in $M_B$, because of obscuration of the central AGN component 
by the dust torus.  On the contrary, no such trend is seen with changing 
radio 
strength.  The $H-K'$ color for radio-quiet AGNs is much the same as 
that for radio-loud AGNs and is in between ``QSOs'' and ``galaxies''.  
This indicates that the radio strength has no significant correlation with 
$M_B$ 
in its range considered here.

In order to see the $z$-dependence,  we show the $J-H$ and $H-K'$ colors 
of intermediate-$z$ AGNs with $0.3<z<0.6$ by dashed lines in Fig. 6 only 
for the 
case of an aperature of 10 pixels.  
It is apparent that the QSO-like colors of brighter AGNs with 
$M_B<-25$ shift bluewards for larger $z$, while galaxies-like colors of 
Seyfert 
1.8 shift redwards. This opposite color trend is consistent with the opposite 
of $k$-corrections between AGNs and galaxies, and calculations based on the 
typical 
SEDs have confirmed that the values of the $k$-corrections for AGNs and 
galaxies 
indeed agree with their respective color shifts as seen in Fig. 6.  
Thus, a criterion of 
$M_B<-27$ is regarded as a discriminator of QSOs, and a criterion of 
Seyfert type 
later than 1.8 as a discriminator of  AGNs dominated the galaxy SED 
component.  
Otherwise, the intermediate colors, as a result of being contributed 
equally from 
QSO and galaxy components, do not show any significant $z$-dependence, 
because 
the $k$-corrections of the different conponents cancel any z-dependence.

%%%%%%%%%%%%%%%%%%%%%%%%%%%%%%%%%%%%%%%%%%%%%%%%%%%%%%%%%%%%%%%%%%%%%%%%%%%%

We are grateful to H. Okuda, M. Narita, and other staff of the infrared 
astronomy group of  the Institute of Space and Astronautical Science 
(ISAS) for their support in using their 1.3m telescope. We thank the 
staff of the Advanced Technology Center of the National Astronomical 
Observatory of Japan (NAOJ) for their new caoting on the mirror 
of the 1.3m telescope at the ISAS.  Gratitude is also extended to the 
Computer Data Analysis Center of the NAOJ.  This work has made use of 
the NASA/IPAC Extra Galactic Database (NED), and has been supported 
partly by the Grand-in-Aid (07CE2002, 10304014) of the Ministry of 
Education, Science, Culture, and Sports of Japan and by the Torey 
Science Foundation.

%%%%%%%%%%%%%%%%%%%
% references
%%%%%%%%%%%%%%%%%%%

%%%%%%%%%%%%%%%%%%%%%%%X
%   figure captions
%%%%%%%%%%%%%%%%%%%%%%%

\clearpage

\figcaption[fig/RA_dec_JHKd.ps]{Distribution of right ascension
     $\alpha$  and declination $\delta$ of 331 AGNs in our sample 
     which were observed on two  different nights ({\it
     filled circles}),  and those observed only once ({\it open
     circles)}.  Top, middle, and bottom panels are for the $J$, 
     $H$,and $K'$ bands, respectively. 
       \label{fig_bunpu_pos}}

\figcaption{
          Distribution of redshift, $z$, and absolute $B$-magnitude, $M_B$,
          of 331 AGNs in our sample which were observed on two
          different nights ({\it filled circles}), and those observed only 
    once ({\it open circles)}.  Top, middle, and bottom panels are for the 
     $J$, $H$,and $K'$ bands, respectively. 
         }

\figcaption{
           Frequency distribution of the transformation error, 
           $\sigma_a$, between
           instrumental and calibrated magnitudes for AGNs.
           Top, middle and bottom panels are for the $J$, $H$, and $K'$ 
           bands, 
           respectively.
         }

\figcaption{
           Frequency distribution of the dispersion,
           $\sigma_d$, in instrumental magnitudes of each standard star
           obtained by dithering.
           The solid line shows the result  based on the data taken in 
           the first period (January 1996$-$April 1996) as well as in the 
           third period (December 1997$-$April 1998), 
           and the dashed line for the second period 
           (November 1996$-$February 1997).
           Top, middle and bottom panels are 
           for the $J$, $H$, and $K'$ bands, 
           respectively
         }

\figcaption{
          Frequency distribution of radio 6 cm flux relative to optical 
          $V$-band flux for AGNs in our sample.
          AGNs with $f_{\nu}({\rm 6cm})/f_{\nu}(V) > 100$ and  
          $f_{\nu}({\rm 6cm})/f_{\nu}(V) < 10$ are classified as radio
          loud and radio quiet, respectively. 
          AGNs in between are classified as intermediate.
          }

\figcaption{NIR $J$-$H$ and $H$-$K'$  colors of AGNs as a function of 
          absolute $B$-magnitude 
          $M_B$, Seyfert type, and radio strength.  Results for low-$z$ 
          AGNs with $0<z<0.3$ 
          are shown by solid lines for the four different apertures of 7, 10, 
          12, and 15 pixels. 
          Results for intermediate-$z$ AGNs with $0.3<z<0.6$ are shown by 
          dashed lines only for an aperture of
          10 pixels.  
          }

%%%%%%%%%%%%%%%%%%%%%%%%
%     figures
%%%%%%%%%%%%%%%%%%%%%%%%

\clearpage

\begin{figure}
  Figure 1\\
  \epsscale{0.8}
  \plotone{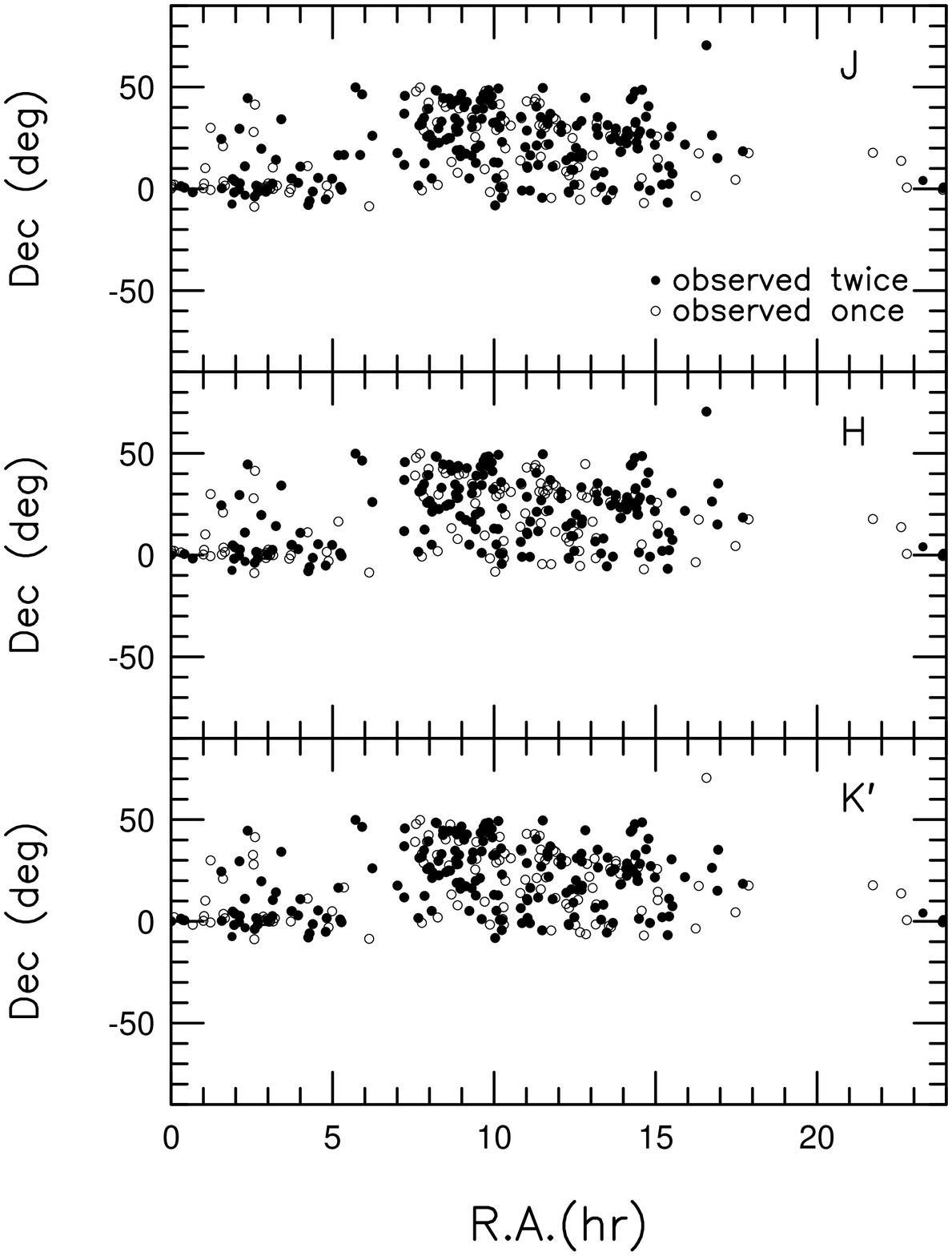}
\end{figure}

\clearpage
\begin{figure}
  Figure 2\\
  \epsscale{0.8}
  \plotone{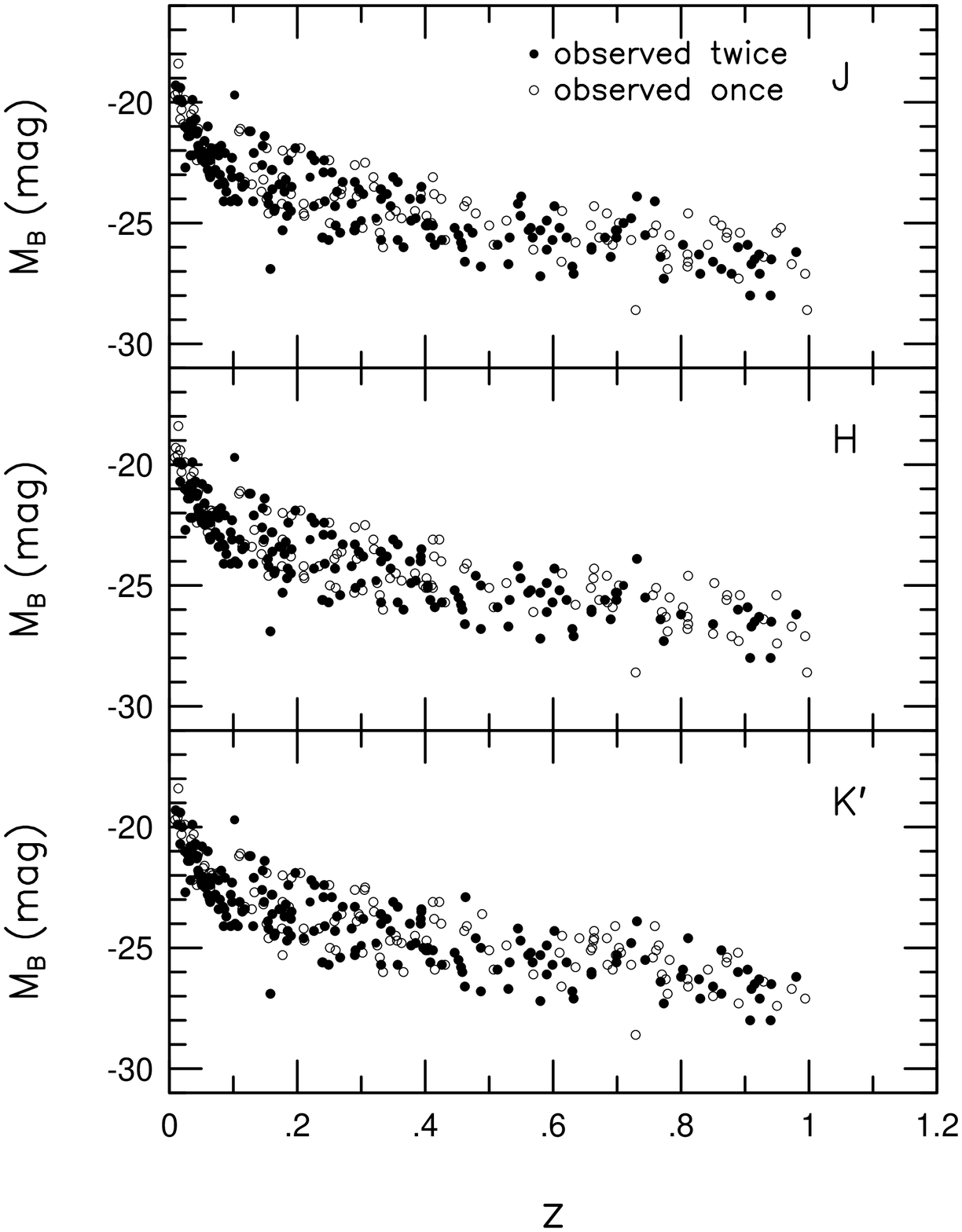}
\end{figure}

\clearpage
\begin{figure}
  Figure 3\\
  \epsscale{0.8}
  \plotone{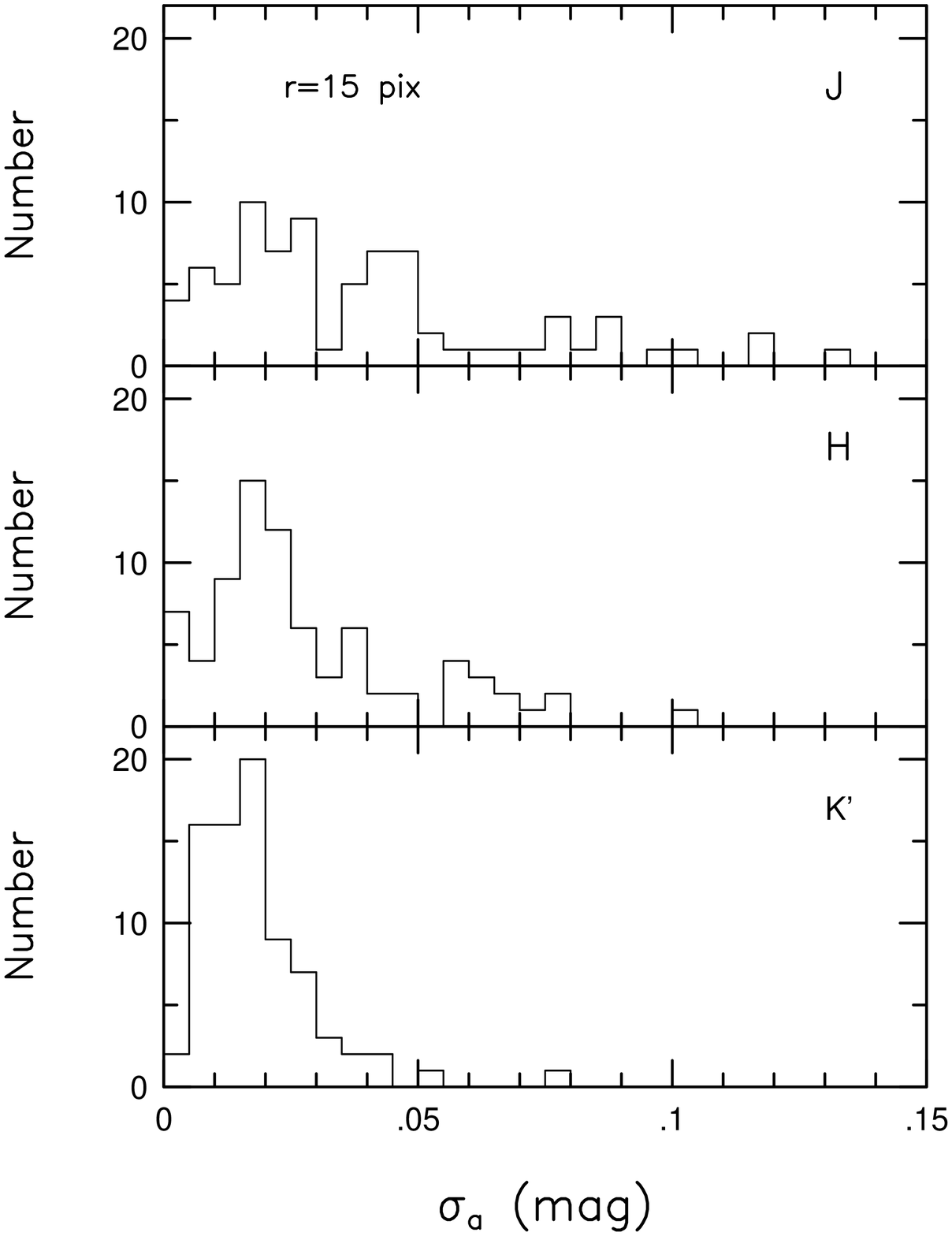}
\end{figure}

\clearpage
\begin{figure}
  Figure 4\\
  \epsscale{0.8}
  \plotone{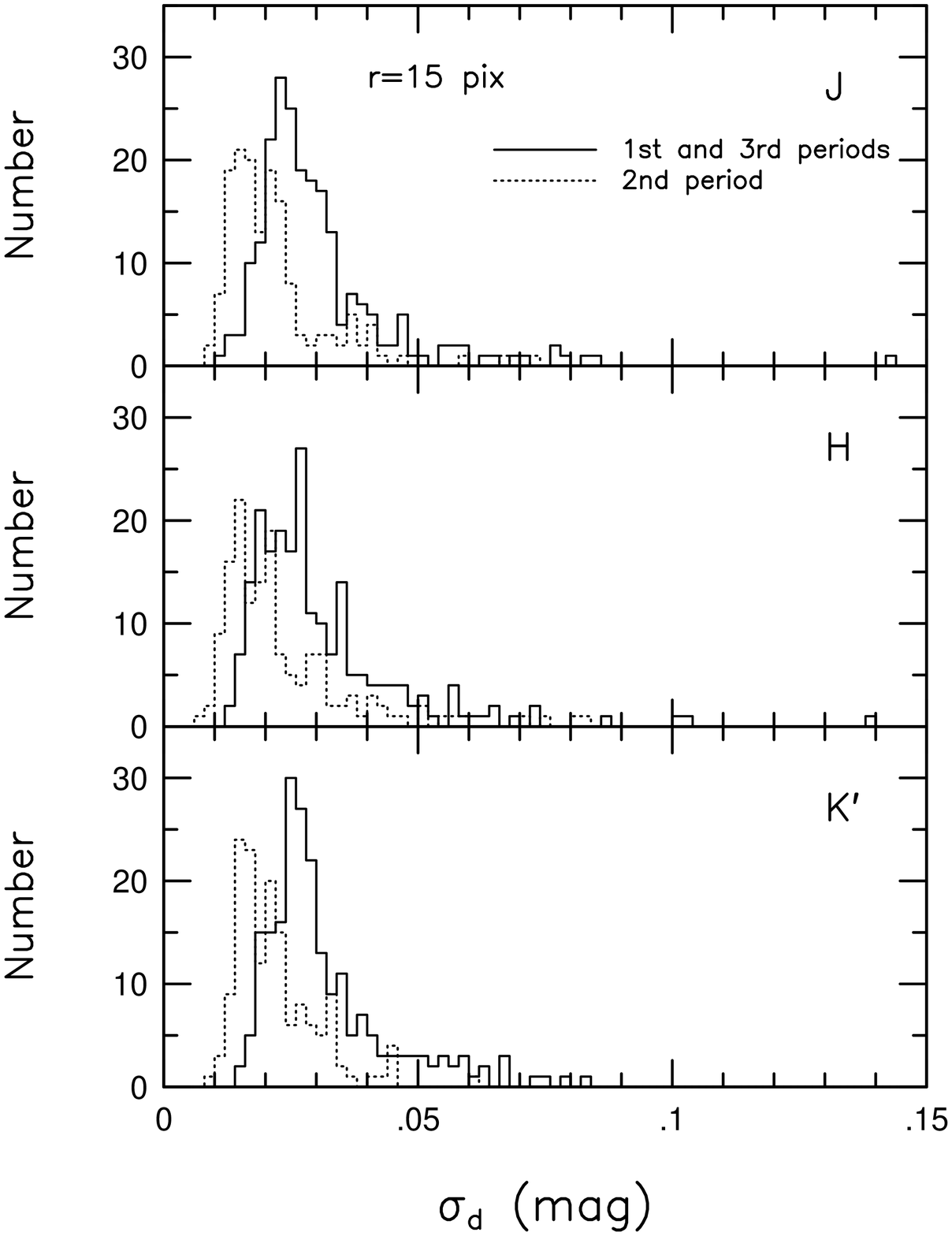}
\end{figure}

\clearpage
\begin{figure}
  Figure 5\
  \epsscale{1.0}
  \plotone{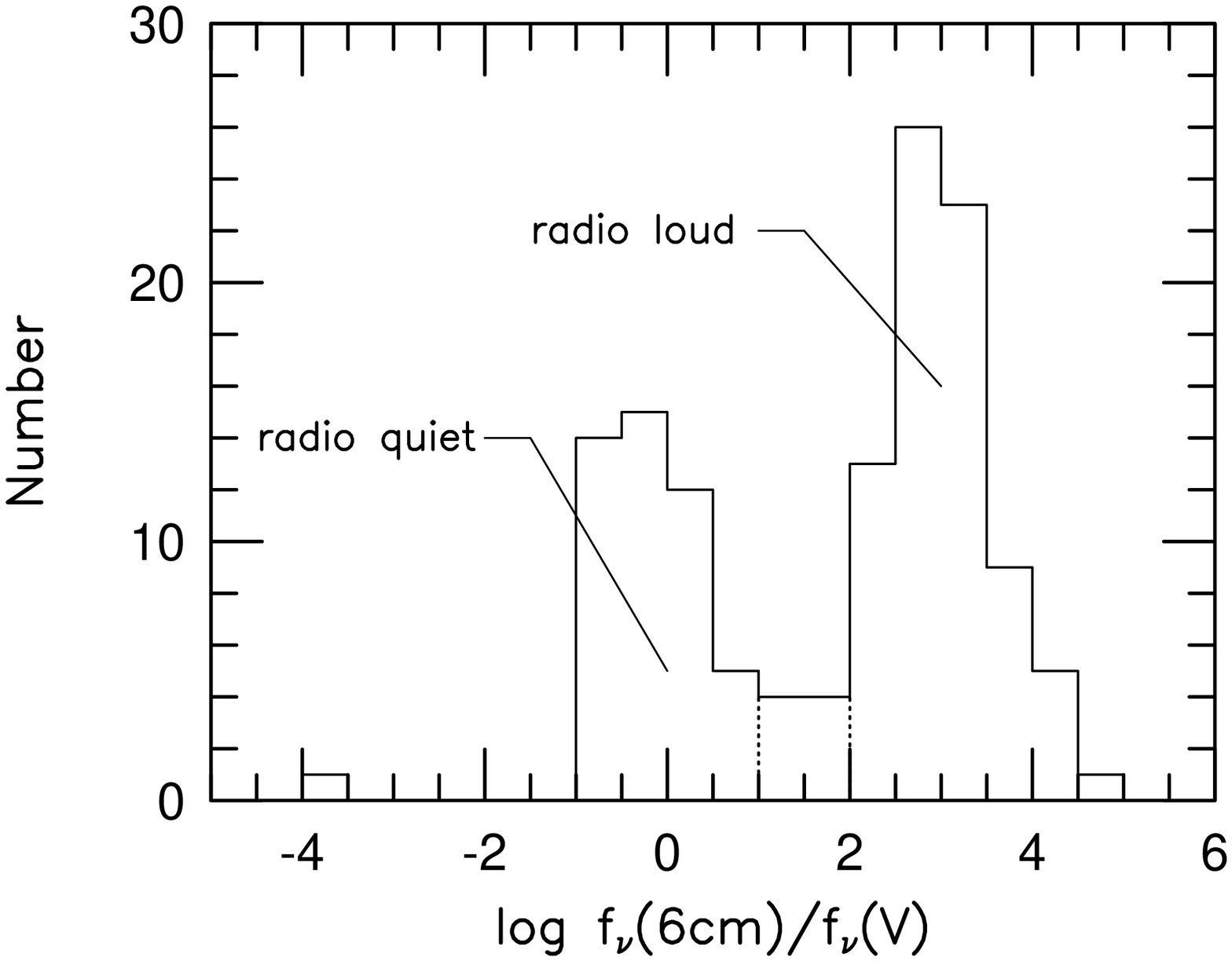}
\end{figure}

\clearpage
\begin{figure}
  Figure 6\
  \plotone{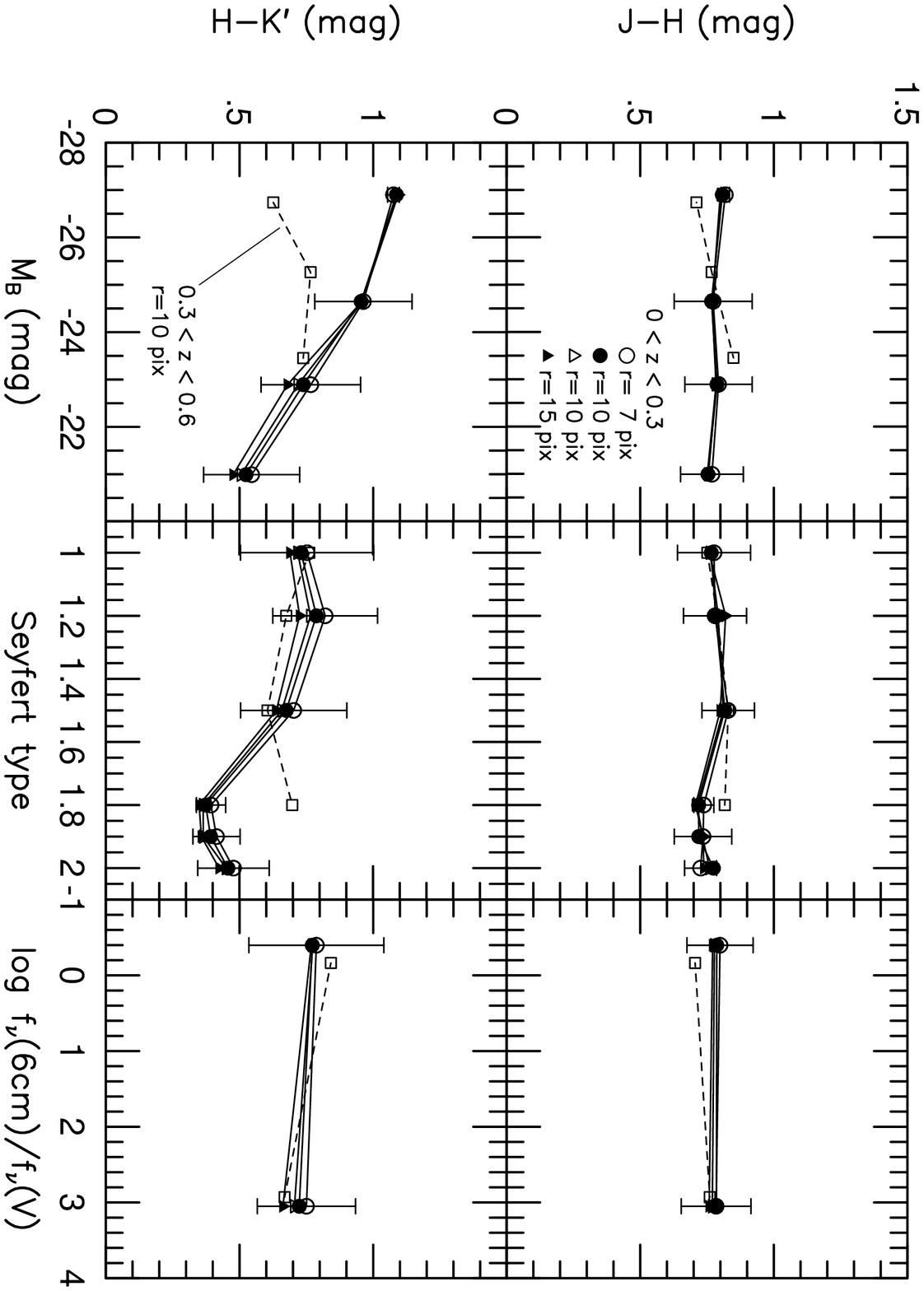}
\end{figure}

%%%%%%%%%%%%%%%%%%%%%%%%%%%%%%%%%%%%%
%   table: list of observed objects  
%%%%%%%%%%%%%%%%%%%%%%%%%%%%%%%%%%%%%

\clearpage



\begin{deluxetable}{clcccccc}
\scriptsize
\tablecaption{Objects list. }
\tablewidth{0pt}
\tablenum{1}
\tablehead{
\colhead{Nmumber} & \colhead{Name}   &
\colhead{$\alpha$(2000)}   & \colhead{$\delta$(2000)} & 
\colhead{$z$}  & \colhead{$M_B$} & \colhead{Seyfert type} 
& \colhead{Loud/Quiet} 
} 
\startdata
  331   &     Q 2352+0025        &   23  54  57.1  &    $+$  00  42  20  &    0.271   &    $-$23.3   &   $-$   &  $-$        \nl
\enddata
\tablecomments{
   Object name, $\alpha(2000)$, $\delta(2000)$, redshift absolute
   $B$-magnitude and Seyfert type are taken from the VV catalog.  
   AGNs with $f_{\nu}({\rm 6cm})/f_{\nu}(V) > 100$ and 
   $f_{\nu}({\rm 6cm})/f_{\nu}(V) < 10$ are classified as radio loud and quiet,
   respectively.  AGNs in between are classified as intermediate.  
   Otherwise the radio 
   6cm flux is not tabulated in the VV catalog.
   }
\end{deluxetable}

%%%%%%%%%%%%%%%%%%%%%%%%%%%%%%%%%%%%%%%%%%%%%%%
\clearpage

\begin{deluxetable}{clcccccccccc}
\scriptsize
\tablecaption{}
\tablewidth{0pt}
\tablenum{2a-Continued}
\tablehead{
  \colhead{Number}   & \colhead{Name}              &
  \colhead{$J_7$}    & \colhead{$\sigma_{J_{7}}$}  &
  \colhead{$J_{10}$} & \colhead{$\sigma_{J_{10}}$} & 
  \colhead{$J_{12}$} & \colhead{$\sigma_{J_{12}}$} &
  \colhead{$J_{15}$} & \colhead{$\sigma_{J_{15}}$} & 
  date               & FWHM$''$        
} 
\startdata
  331   &       Q 2352+0025        &   15.58   &    0.05   &   15.52   &    0.06   &   15.46   &    0.07   &   15.47    &    0.09    &      96/12/23   &    4.1   \nl 
        &                          &   15.60   &    0.06   &   15.51   &    0.09   &   15.56   &    0.11   &   15.71    &    0.18    &      97/12/31   &    3.8   \nl 
\enddata
\tablecomments{The number in the subscript corresponds to the 
   aperture radius in unit of pixel. The pixel scale in this
   study is 1.07$''$pix$^{-1}$.
}
\end{deluxetable}

%%%%%%%%%%%%%%%%%%%%%%%%%%%%%%%%%%%%%%%%%%%%%%%
\clearpage

\begin{deluxetable}{clcccccccccc}
\scriptsize
\tablecaption{}
\tablewidth{0pt}
\tablenum{2b-Continued}
\tablehead{
  \colhead{Number}   & \colhead{Name}              &
  \colhead{$H_7$}    & \colhead{$\sigma_{H_{7}}$}  &
  \colhead{$H_{10}$} & \colhead{$\sigma_{H_{10}}$} & 
  \colhead{$H_{12}$} & \colhead{$\sigma_{H_{12}}$} &
  \colhead{$H_{15}$} & \colhead{$\sigma_{H_{15}}$} & 
  date               & FWHM$''$        
} 
\startdata
  331   &       Q 2352+0025        &   14.60   &    0.06   &   14.60   &    0.09   &   14.57   &    0.11   &   14.55    &    0.15    &      96/12/23   &    3.9   \nl 
        &                          &   14.56   &    0.07   &   14.49   &    0.09   &   14.41   &    0.10   &   14.27    &    0.12    &      97/12/31   &    3.6   \nl 
\enddata
\tablecomments{The number in the subscript corresponds to the 
   aperture radius in unit of pixel. The pixel scale in this
   study is 1.07$''$pix$^{-1}$.
}
\end{deluxetable}

%%%%%%%%%%%%%%%%%%%%%%%%%%%%%%%%%%%%%%%%%%%%%%%
\clearpage

\begin{deluxetable}{clcccccccccc}
\scriptsize
\tablecaption{}
\tablewidth{0pt}
\tablenum{2c-Continued}
\tablehead{
  \colhead{Number}   & \colhead{Name}              &
  \colhead{$K'_7$}    & \colhead{$\sigma_{K'_{7}}$}  &
  \colhead{$K'_{10}$} & \colhead{$\sigma_{K'_{10}}$} & 
  \colhead{$K'_{12}$} & \colhead{$\sigma_{K'_{12}}$} &
  \colhead{$K'_{15}$} & \colhead{$\sigma_{K'_{15}}$} & 
  date               & FWHM$''$        
} 
\startdata

  331   &       Q 2352+0025        &   13.65   &    0.05   &   13.50   &    0.05   &   13.48   &    0.06   &   13.40    &    0.08    &      96/12/23   &    3.9   \nl 
        &                          &   13.80   &    0.05   &   13.80   &    0.07   &   13.72   &    0.08   &   13.65    &    0.09    &      97/12/31   &    3.5   \nl 
\enddata
\tablecomments{The number in the subscript corresponds to the 
   aperture radius in unit of pixel. The pixel scale in this
   study is 1.07$''$pix$^{-1}$.
}
\end{deluxetable}

%%%%%%%%%%%%%%%%%%%%%%%%%%%%%%%%%%%%%%%%%%%%%%%%%%%%%%%%%%%%%%%%%%%%%%%%%%%%%
\end{document}